\documentclass[]{pasj01}

\usepackage{natbib}
\usepackage{color}

\usepackage{ulem}

\usepackage{bm}

\newcommand{\Msun}{\, M_{\odot}}
\newcommand{\Rsun}{R_{\odot}}

\def\gsim{\mathrel{\rlap{\lower 4pt \hbox{\hskip 1pt $\sim$}}\raise 1pt
\hbox {$>$}}}
\def\lsim{\mathrel{\rlap{\lower 4pt \hbox{\hskip 1pt $\sim$}}\raise 1pt
\hbox {$<$}}}

\newcommand{\vns}{v_{\rm NS}}

\begin{document}

\title{

Long-term Simulations of Multi-Dimensional Core-collapse Supernovae: Implications for Neutron Star Kicks}

\author{Ko \textsc{Nakamura}\altaffilmark{1,2*}
}
\altaffiltext{1}{Department of Applied Physics, Fukuoka University, Nanakuma 8-19-1, Johnan, Fukuoka 814-0180, Japan}
\altaffiltext{2}{Research Insitute of Stellar Explosive Phenomena, Fukuoka University, Nanakuma 8-19-1, Johnan, Fukuoka 814-0180, Japan}
\email{nakamurako@fukuoka-u.ac.jp}

\author{Tomoya \textsc{Takiwaki},\altaffilmark{3}}
\altaffiltext{3}{Division of Theoretical Astronomy, National Astronomical Observatory of Japan, 2-21-1, Mitaka, Tokyo 181-8588, Japan}

\author{Kei \textsc{Kotake}\altaffilmark{1,2}}


\maketitle

\begin{abstract}
  Core-collapse supernovae (CCSNe) are the final stage of massive stars, marking the birth of neutron stars (NSs). The aspherical mass ejection drives a natal kick of the forming NS. 
  In this work we study the properties of the NS kick
  based on our long-term hydrodynamics CCSN simulations.
  We perform two-dimensional (2D) simulations for ten progenitors
  from a 10.8 to 20 $\Msun$ star covering a wide range of the progenitor's compactness parameter, and two three-dimensional (3D) simulations for an
  11.2 $\Msun$ star. 
  Our 2D models present a variety of explosion energies between
  $\sim 1.3 \times 10^{50}$ erg and $\sim 1.2 \times 10^{51}$ erg, and
  NS kick velocities between
  $\sim 100$ km s$^{-1}$ and $\sim 1500$ km s$^{-1}$.
  For the 2D exploding models, we find that the kick
  velocities tend to become higher with the progenitor's
compactness. This is because the high progenitor compactness results in high neutrino luminosity from the proto-neutron star (PNS),
leading to more energetic explosions.
Since high-compactness progenitors produce
massive PNSs, we point out that the NS masses
and the kick velocities can be correlated, which is
moderately supported by observation. 
Comparing 2D and 3D models of the 11.2 $\Msun$ star, 
the diagnostic explosion energy in 3D is, as 
previously identified, higher than that in 2D, whereas the 3D model results in a smaller asymmetry in the ejecta distribution and 
a smaller kick velocity than in 2D. Our results confirm the importance of self-consistent CCSN modeling covering a long-term postbounce evolution in 3D for a quantitative prediction of the NS kicks.
\end{abstract}

\section{Introduction}
\label{sec:intro}

Young radio pulsars have been measured to possess average
velocities as high as 200-500 km s$^{-1} $ 
\citep[e.g.,][]{lyne,kaspi,arzoumanian,hobbs}.
Some of them have kick velocities higher than 500 km s$^{-1}$, 
such as the compact remnant RX J0822-4300 in Puppies A and PSR B1508+55 
(e.g., \cite{katsuda18} for collective references therein). 
It has long been proposed that these high
kick velocities could be produced by asymmetric mass
 ejection when a
 core-collapse supernova (CCSN) explosion is initiated in a non-spherical manner 
\citep[``hydrodynamic kick scenario'', e.g.,][]{janka94, burrows96} 
or by anisotropic neutrino emission from the proto-neutron star 
\citep[PNS; ``neutrino-induced kick scenario'', e.g.,][]{woosley87,bisnovatyi93,lai,kotake05,fryer06,kusenko08,sagert08}.

 In the hydrodynamic kick scenario, the non-radial instabilities
 (convective overturn and the standing accretion shock instability, SASI:
 \citep{blondin03,foglizzo06,foglizzo07})
 play a key role in producing mass ejection asymmtries during the onset of 
   the neutrino-driven explosion (see \citet{janka16} for a review). This has been demonstrated by
 two-dimensional (2D) and three-dimensional (3D) hydrodynamics
 simulations showing that the kick velocities can be as high as
 $\sim 1000$ km s$^{-1}$
 \citep[e.g.,][]{scheck04,scheck06,nordhaus,nordhaus12,wongwathanarat13,janka17,mueller17}.
 The kick velocities are generally imparted opposite to the direction
 of the stronger explosion, where the explosively nucleosynthesized
 elements are preferentially expelled \citep{wongwathanarat13}. Based on
 three-dimensional simulations,
 \citet{wongwathanarat13} were the first
  to propose that the forming NS is accelerated via the
  "gravitational tug-boat mechanism," whereby 
  the asymmetric ejecta exerts a long-lasting momentum transfer to the NS
   by the gravitational pull over a period of seconds.

 In fact, recent systematic measurements of X-ray morphologies
 for Galactic supernova remnants (SNRs) support
  the hydrodynamic kick scenario. They present evidence that the bulk of the supernova ejecta moves
 in the opposite directions to the proper motion of the NSs
 \citep{holland17, bear18, katsuda18}.  For example, detailed X-ray mapping of Cas A and G292.0+1.8 has revealed that the bulk motion of the total ejecta is roughly in the opposite direction to the apparent motion of the NS.
 In the Puppies A SNR, optical fast-moving oxygen-rich knots were
 observed in the opposite direction to the proper motion of the NS. 
 No correlation was observed
 between the kick velocities and the
 magnetic field strengths of the NSs \citep{katsuda18}, which conflicts
 with the neutrino-induced kick scenario assuming extremely
 strong NS magnetic fields ($\geq 10^{16}$ G).

 In the seminal works by \citet{scheck04,scheck06} and \citet{wongwathanarat13},
 it was shown that
 the kick velocities are connected to explosion properties such as
 the explosion energy and the mass of the ejected material and the central remnant.
 In order to clarify the connection between the explosion properties and
  the progenitor structure,
 one needs self-consistent supernova models. In the above studies,
 limited progenitors were employed ($15~\Msun$ and $20~\Msun$), where
 the central core was excised to follow the long-term evolution (see also
  \citet{gessner}). In the context
 of self-consistent simulations
 \citep{nordhaus12,bruenn13,muellerb15,pan,summa16,mueller17,vartanyan,oconnor18},
 the progenitor dependence on the NS kick has not yet been studied.
 It should be mentioned that \citet{janka17} investigated how the neutron star
 kick depends on the energy, ejecta mass, and asymmetry of the
 supernova explosion based on an analytical scaling relation (see also
 \citet{bray16,bray18}). These estimates should be validated by outcomes
 from self-consistent simulations.

Very recently \citet{bernhard19} presented 3D CCSN simulations of low-mass (low-compactness) progenitors until 1--2 s after bounce 
and proposed a possible correlation between kick velocity and some explosion properties.
In this paper we present results of 2D CCSN simulations of
$10.8~\Msun$--$20.0~\Msun$ stars of \citet{WHW02}
for longer-term post-bounce evolution covering a wide range of the progenitor's compactness parameter. 
We pay attention to the progenitor's compactness parameter $\xi$
\citep{oconnor11} because it has been shown as one of the key parameters 
for diagnosing the explosion properties in both 1D \citep{oconnor11,oconnor13,ugliano12,ertl16,sukhbold16} and 2D models \citep{nakamura15,summa16}.
In order to obtain a saturated value of the
 explosion energy, we had to
follow a long-term evolution up to $\sim$ 7 s after bounce 
of a $17 M_{\odot}$ star
in 2D (e.g., \citet{nakamura16}). Given the
 computational cost, we employ the similar numerical scheme to 
\citet{nakamura15} where the isotropic diffusion source approximation (IDSA,
 \citet{idsa})
is used for spectral transport of
electron and anti-electron neutrinos and a leakage scheme is employed for
heavy lepton neutrinos\footnote{Note that our updated code \citep{kotake18} that can deal with three
  flavor neutrino transport with more detailed neutrino opacities
  is unfortunately too computational expensive for the purpose of this work.}.
Since a large explosion asymmetry has been seen in recent 3D CCSN models \citep{hanke13,takiwaki14,ernazar15,lentz15,melson15,takiwaki16,roberts16,bernhard17,kuroda17,takiwaki18,oconnor18_3d,ott18,glas18,vartanyan19,burrows19},
we also perform a 3D simulation for an
$11.2~\Msun$ progenitor star (this progenitor model was also employed in \citet{muellerb15}). We
compare the kick properties with those from the corresponding 2D model. 

 This paper is organized as follows. Section \ref{sec:method} summarizes
 our numerical methods and the progenitor models employed in this work.
We present our 2D and 3D results in Sections \ref{sec:2d} and \ref{sec:3d}, respectively.
We conclude with discussions in Section \ref{sec:discuss}. 
In the Appendix \ref{sec:accr}
we present a caveat for the 2D models. 
Unless otherwise stated, time is measured after bounce throughout this paper.

\section{Method}
\label{sec:method}

We employ ten progenitors from \citet{WHW02}
covering zero-age main-sequence (ZAMS) masses from $10.8 \, \Msun$
to $20.0 \, \Msun$. 
All of the progenitors were trending towards an explosion
in our previous 2D simulations covering $\sim 1$ s after bounce \citep{nakamura15}. 
In this paper we show the shock evolution up to  
a later post-bounce time ($\sim 8$ s) in 2D.
One of the progenitors is also simulated in 3D. 
Given that the progenitor's compactness parameter $\xi_M$, which is defined in  \citet{oconnor11} as a function of an enclosed mass $M$, 
\begin{equation}
\xi_M = \frac{M/\Msun}{R(M)/1000 \, {\rm km}},
\end{equation}
is a good diagnostics for the explosion properties.  The ten progenitors are taken to cover low to high $\xi_M$ among the exploding models. The progenitor properties are summarized in Table \ref{t1}. 
In this paper we estimate the compactness parameter $\xi_M$ at $M=2.5 \, \Msun$ from the progenitor models. 

The Newtonian hydrodynamics code that we employ in this work is essentially the same as that in \citet{nakamura15} except for some minor revisions. 
To follow a long-term evolution, the spatial range of the computational domain
 is extended from 5,000 km in radius to 100,000 km in this 2D study. 
 The outer boundaries of all the models examined
 are located in the carbon-helium layers.
 The computational domain is sufficiently large to prevent the SN shock from being
 affected by the boundary condition before shock revival. 
The 2D models are computed on a spherical coordinate grid 
with a resolution of $n_r \times n_\theta = 1008 \times 128$ zones. 
For 3D models, we put the outer boundary at 10,000 km and simulate with the resolution of $n_r \times n_\theta \times n_\phi = 648 \times 64\times 128$ zones. 
Our spatial grid has a finest mesh spacing $dr_{\rm min} = 250$ m at the center, 
and $dr/r$ is better than 1.0 \% at $r > 100$ km.
Seed perturbations for aspherical instabilities are imposed by hand at 10 ms after bounce by
introducing random perturbations of $0.1\%$ in density on the whole
computational grid except for the unshocked core.

For electron and anti-electron neutrinos, we employ the isotropic diffusion
source approximation \citep[IDSA,][]{idsa}, taking 20 energy bins with an upper bound of 300 MeV.
For heavy-lepton neutrinos a leakage scheme is employed (see \citet{nakamura15} for more detail). Regarding the equation of state (EOS), 
we use that of \citet{lattimer91} 
with a nuclear incomprehensibility of $K = 220$ MeV.
At low densities, 
we employ an EOS accounting for photons, electrons, positrons, 
and ideal gas contribution. 
We follow the explosive nucleosynthesis 
by solving a simple nuclear network consisting of 13 alpha-nuclei: 
$^4$He, $^{12}$C, $^{16}$O, $^{20}$Ne, $^{24}$Mg, $^{28}$Si, 
$^{32}$S, $^{36}$Ar, $^{40}$Ca, $^{44}$Ti, $^{48}$Cr, $^{52}$Fe, and $^{56}$Ni.
Feedback from the composition change to the EOS is neglected, 
whereas the energy feedback from the nuclear reactions 
to the hydrodynamic evolution is taken into account as in \citet{nakamura14a}.

Note that our simulations exploit Newtonian gravity, which is inadequate for accurate modeling of CCSNe. 
The main goal of this series of systematic CCSN study \citep{nakamura15,Horiuchi17,Horiuchi18} is to explore qualitative properties of CCSNe using self-consistent models. 
More sophisticated CCSN modeling with general relativistic effects is necessary for an accurate prediction of CCSN properties, including the NS kick.
Our 3D CCSN simulations taking account of general relativistic effects will be reported in a forthcoming paper.

\begin{table*}[htb]
\begin{center}
\caption{Summary of the initial models in this work. All the progenitor models are taken from \citet{WHW02}. Model name denotes solar metallicity (s) and zero-age main sequence mass in units of solar mass. Progenitor mass and radius, as well as Fe core mass and radius ($M_{\rm Fe}$, $R_{\rm Fe}$), the interface radius between CO and HeC layers ($R_{\rm CO/HeC}$), and the mass included in the computational domain ($M_{\rm comp}$) are listed. The compactness parameter ($\xi_{2.5}$) is estimated at $M=2.5 \Msun$ from the pre-collapse progenitor data.}\label{t1}
\begin{tabular}{cccccccc}
\hline 
Progenitor & Mass & Radius & $M_{\rm Fe}$ & $R_{\rm Fe}$ & 
                $R_{\rm CO/HeC}$ & $M_{\rm comp}$ & $\xi_{2.5}$\\
           & ($\Msun$) & ($\Rsun$) & ($\Msun$) & (km) &
               (km) &  ($\Msun$) & \\
\hline 
s10.8 & 10.4 & 563 & 1.36 & 1560 & 17800 & 1.82 & 0.003\\ 
s11.0 & 10.6 & 587 & 1.37 & 1460 & 25400 & 1.87 & 0.004\\ 
s11.2 & 10.8 & 596 & 1.25 & 1000 & 33500 & 1.91 & 0.005\\ 
s12.4 & 11.0 & 680 & 1.45 & 1590 & 34500 & 2.55 & 0.028\\ 
s13.8 & 11.8 & 774 & 1.48 & 1590 & 40600 & 3.03 & 0.081\\ 
s16.0 & 13.2 & 913 & 1.44 & 1580 & 50900 & 3.69 & 0.154\\
s17.0 & 13.8 & 958 & 1.44 & 1500 & 54400 & 4.06 & 0.161\\
s19.6 & 13.4 & 1160 & 1.47 & 1570 & 88600 & 5.04 & 0.119\\ 
s19.8 & 14.5 & 1130 & 1.44 & 1500 & 80700 & 5.02 & 0.136\\ 
s20.0 & 14.7 & 1120 & 1.46 & 1690 & 84200 & 5.10 & 0.127\\
\hline 
\end{tabular}
\end{center}
\end{table*}

\section{Results from 2D simulations}
\label{sec:2d}

In this section we estimate the kick velocities of the (forming) NSs in our 2D models and attempt to compare them with observations. As is well known, the 2D assumption leads to an artificially powerful kick along the symmetry axis, making a quantitative comparison between models and observations difficult. However, we start from 2D models because qualitative discussions  about the progenitor dependence of the kick velocities is yet to be explored even in (the context of self-consistent) 2D modeling, which we think still meaningful.

\subsection{Neutron star kick}
\label{sec:kick}
Our numerical code (grid setup) does not conserve the momentum and is unable to directly observe the central NS motion. 
More sophisticated methods employed by \citet{nordhaus} and \citet{nagakura17} make it possible. 
A simple method to estimate NS kick velocity is to sum the net momentum of ejecta and convert it into a recoil velocity assuming momentum conservation.
Another way to evaluate the kick velocity in such a momentum-non-conserving simulation is to integrate forces acting on the PNS. 
The kick velocity evaluated by force integration is, however, not so different from the recoil velocity given by a simple formula \citep{wongwathanarat13}. 
Therefore, we use the simple recoil formula \citep{scheck04} in the current paper. 

Following \citet{scheck04}, we estimate the asymmetry of the ejected matter by $\alpha_{\rm gas}$,
\begin{equation}\label{eq:2}
\alpha_{\rm gas} \equiv | P_{z, {\rm gas}} | / P_{\rm gas} \equiv | \int {\rm d}m \,v_z | /  \int {\rm d}m \,|\vec{v}|,
\end{equation}
where $P_{z, {\rm gas}}$ is the $z$-component (along the 2D axis) of the total momentum ($P_{\rm gas}$) of the ejecta, $v_z$ is the fluid velocity ($v$) along the axis, and $m$ denotes the mass coordinates. The integrals are performed over the ``ejecta'' mass with the positive local energy and positive radial velocity at each time. 

The kick velocity, $\vns$, is estimated using $\alpha_{\rm gas}$ as 
\begin{equation}\label{eq:3}
v_{\rm NS} = \alpha_{\rm gas} P_{\rm gas} / M_{\rm NS},
\end{equation}
 where the NS surface is defined at a fiducial density of $\rho = 10^{11} \,{\rm g \, cm}^{-3}$ and the 
 baryonic 
 NS mass ($M_{\rm NS}$) is estimated by the enclosed mass.

\begin{figure}[h]
\begin{center}
\includegraphics[width=0.95\linewidth]{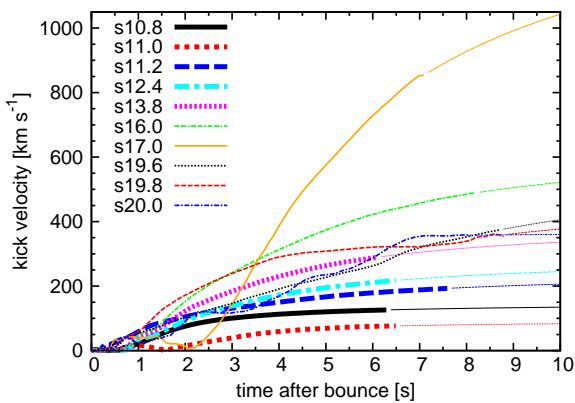}
\end{center}
\caption{
Time evolution of the kick velocity in the 2D models.
At the final time of our simulations, the kick velocity is mostly in the range $\sim$ 200 -- 400 ${\rm km \, s^{-1}}$, whereas it has a wide variety between $\sim 100 \, {\rm km \, s^{-1}}$ (slowest one for model s11.0, thick red dotted line) and $\sim 850 \, {\rm km \, s^{-1}}$ (fastest one for model s17.0, orange solid line) at the time each simulation terminates. 
We use Equation (\ref{eq:extra}) and extrapolate the kick beyond the final simulation time (thin lines).
}\label{f1}
\end{figure}

Figure \ref{f1} shows the time evolution of $\vns$ for all the 2D models in this work. 
It shows a wide variety from $v_{\rm NS} \sim 100$ km s$^{-1}$ (model s11.0) to $\sim 850$ km s$^{-1}$ (model s17.0) at the final simulation time. Note that the kick velocities from most of the 2D models are as fast as young radio pulsars (several hundred km s$^{-1}$). For models s11.0 and s17.0, the kick velocity ($\vns$) temporarily becomes zero a few seconds after bounce. 
This is because the shock expansion of these models first occurs along one direction then transits to other direction (which we will discuss with reference to Figure \ref{f3}). The shock revival occurs earliest 
 for model s11.2 (thick blue dashed line) among the 2D models, which shows the earliest rise in $\vns$; however, the final value is relatively small among the models.

\begin{figure*}[htb]
\begin{center}
\begin{tabular}{cc}
\includegraphics[width=0.5\linewidth]{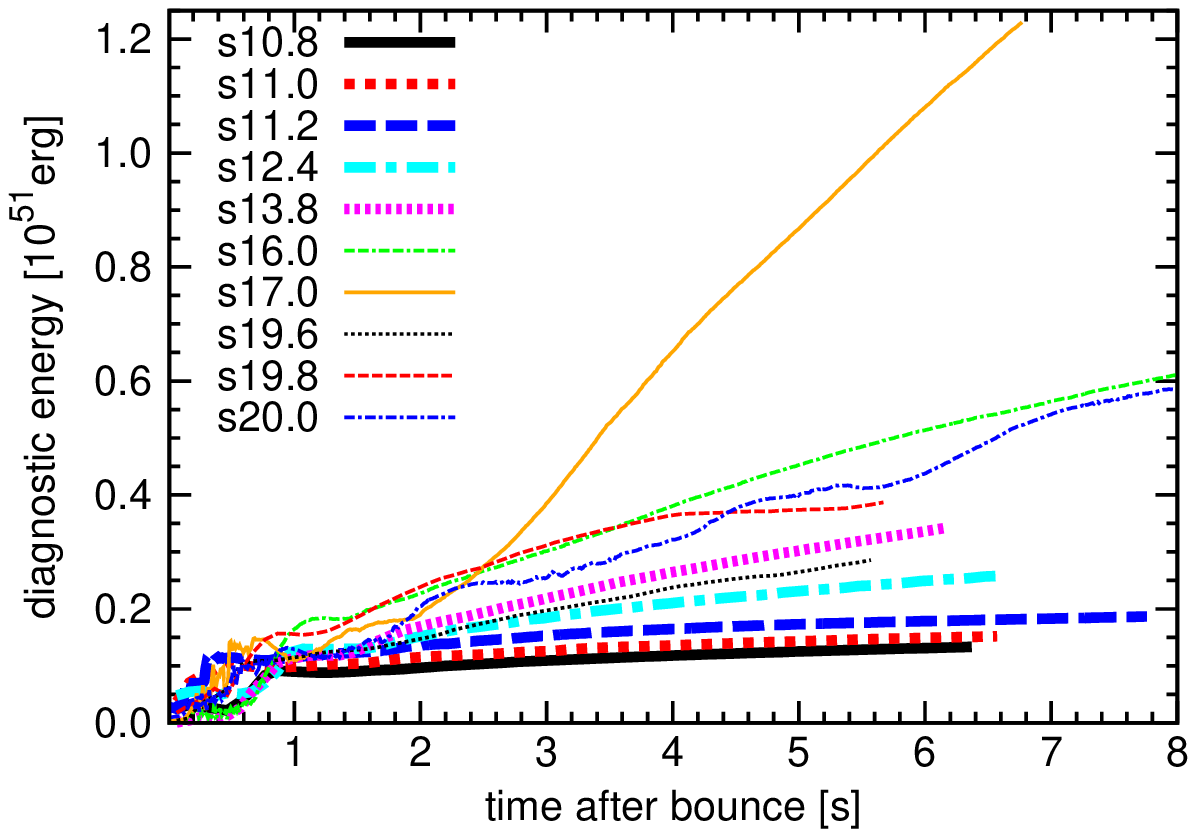} &
\includegraphics[width=0.5\linewidth]{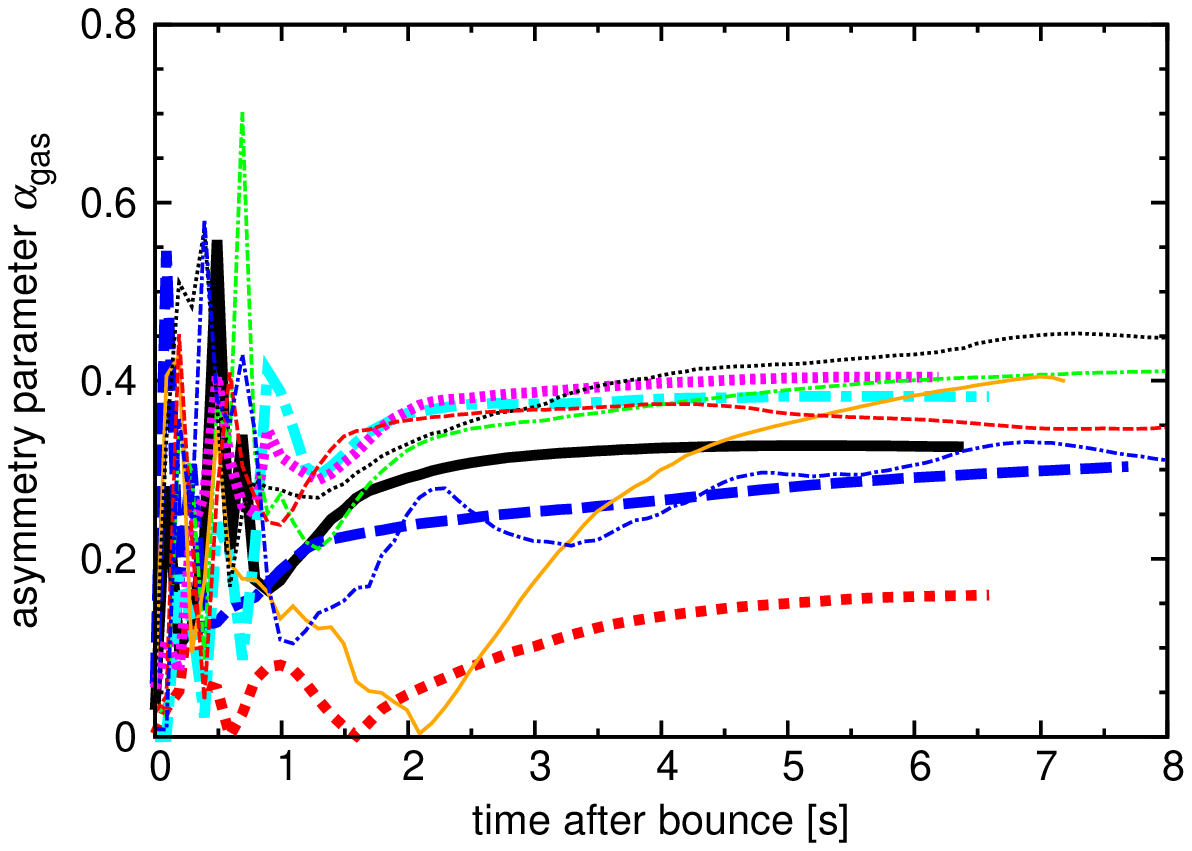} 
\end{tabular}
\end{center}
\caption{Time evolutions of the diagnostic explosion energy (left panel) and the asymmetry parameter of the ejecta ($\alpha_{\rm gas}$, right panel), respectively. Note that the same line styles are used in both of the panels.}
\label{f2}
\end{figure*}

\begin{figure*}[htb]
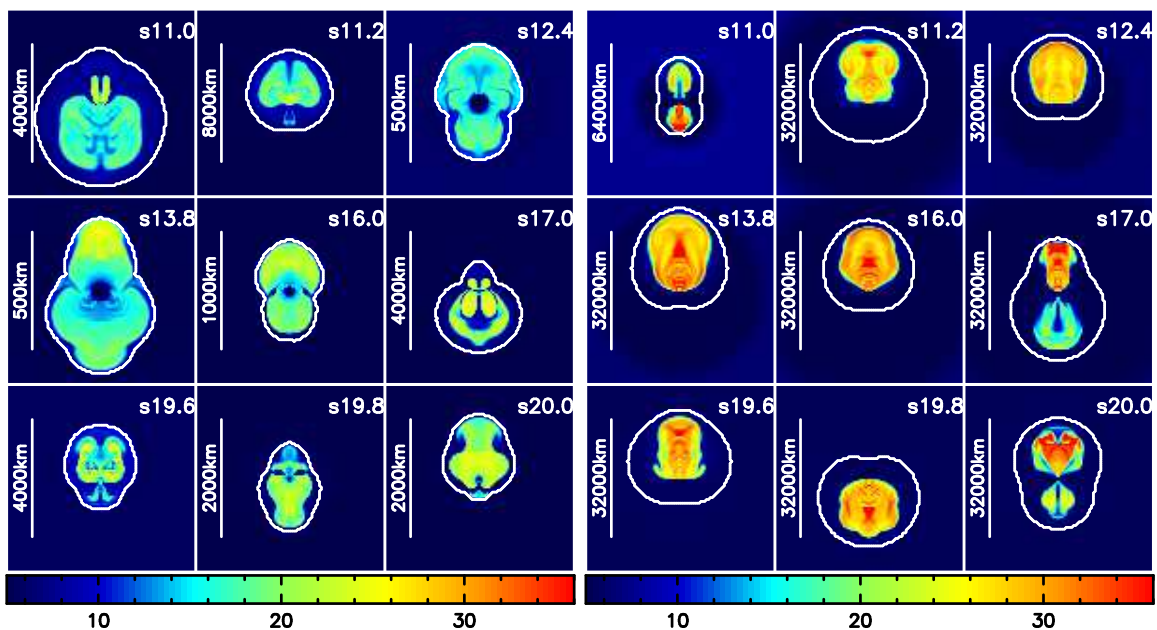

\begin{center}
\begin{tabular}{cc}
\includegraphics[width=0.48\textwidth,angle=270]{pgp1-0500low.eps}
\includegraphics[width=0.48\textwidth,angle=270]{pgp1-2501low.eps}
\end{tabular}
\end{center}
\caption{Snapshots of the specific entropy $s$ in $k_{\rm B}$/baryon for nine models at 0.5 s (left panels) and 2.5 s (right panels) after bounce, respectively. The model name is denoted in the top right corner of each panel. Note the different spacial scale in each panel, which is indicated by the vertical scale bar. White lines surrounding the high-entropy region represent the position of shock waves. In the left panels, most of the models present a shock expansion, except for models s12.4 and s13.8 which still show a shock sloshing within $\sim 250$ km from the center. The right panels show that all of the models have a maximum shock radius larger than 10,000 km. }
\label{f3}
\end{figure*}

The $\vns$ estimated from Equation (\ref{eq:3}) is determined by $\alpha_{\rm gas}$, $P_{\rm gas}$, and $M_{\rm NS}$. 
The total momentum of ejecta $P_{\rm gas}$ is tightly connected to the vigorousness of explosion, or the explosion energy. 
The diagnostic explosion energy of our 2D models is shown in the left panel of Figure \ref{f2}. 
Here, the diagnostic energy is defined as the sum of the total energy (kinetic, internal and gravitational energies) in the ``ejecta'' region where the total energy and radial velocity are positive. 
Comparing with Figure \ref{f1},  one can see that 
the diagnostic energies show a similar evolution to the kick velocity, which implies that $P_{\rm gas}$ is the dominant factor in the estimation of $\vns$ in Equation (\ref{eq:3}).

On the other hand, the asymmetry parameter $\alpha_{\rm gas}$ does not show such a clear correlation to $\vns$. 
For example, the asymmetry parameter of model s11.2 (thick blue dotted line) is almost converged at $\alpha_{\rm gas} \sim 0.3$, which is very close to that of model s20.0 (thin blue dash-dotted line), whereas the kick velocity of model s11.2 is much smaller than that of model s20.0 ($350 \, \rm km \, s^{-1}$, Figure \ref{f1}).
This is caused by the less energetic explosion of model s11.2, which results in a smaller value of the integrated gas momentum $P_{\rm gas}$. 

It should be noted that the models showing a 
unipolar explosion result in higher $\alpha_{\rm gas}$ than those showing a bipolar explosion. 
Figure \ref{f3} depicts the evolution of the blast geometry for representative 2D models 
at 500 ms (left panels) and 2.5 s (right panels) 
after bounce, respectively.
The shock of model s13.8 (middle-left panel in each plot of Figure \ref{f3}), for example, is expanding to the northern direction, which results in high $\alpha_{\rm gas}$ (thick magenta dotted line in Figure \ref{f2}). 
On the other hand, bipolar explosion models (s11.0, s11.2, s17.0 and s20.0, see right panels of Figure \ref{f3}) have smaller $\alpha_{\rm gas}$ (0.1--0.25) compared with the other models ($\alpha_{\rm gas} > 0.3$) at this time.
Some of the models, s11.0 and s17.0, 
change the direction of the shock expansion during the long-term evolution. This leads to a significant change in the kick velocity as already mentioned (e.g., the zero-crossing feature in $\vns$ (Figure \ref{f1}) as well as in $\alpha_{\rm gas}$ (right panel of Figure \ref{f2})). 

\begin{figure*}[htb]
\begin{center}
\begin{tabular}{cc}
\includegraphics[width=0.5\linewidth]{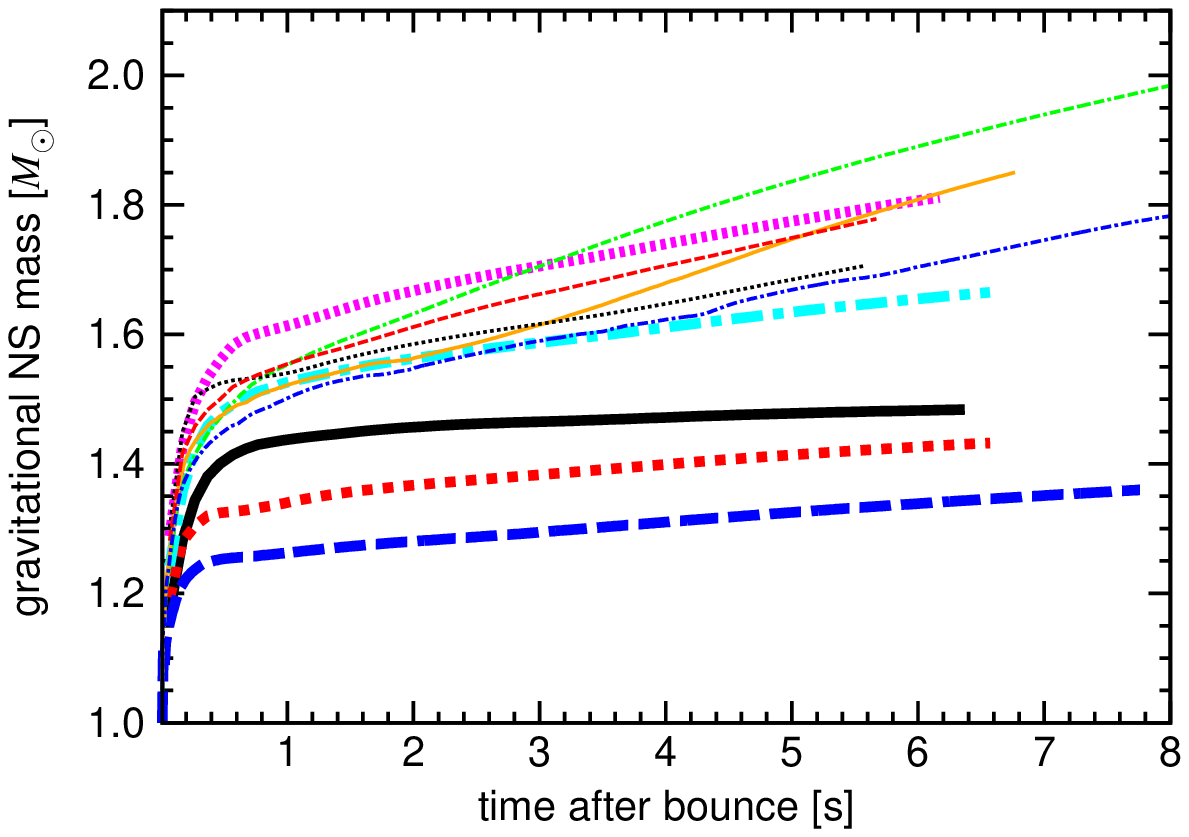}  &
\includegraphics[width=0.5\linewidth]{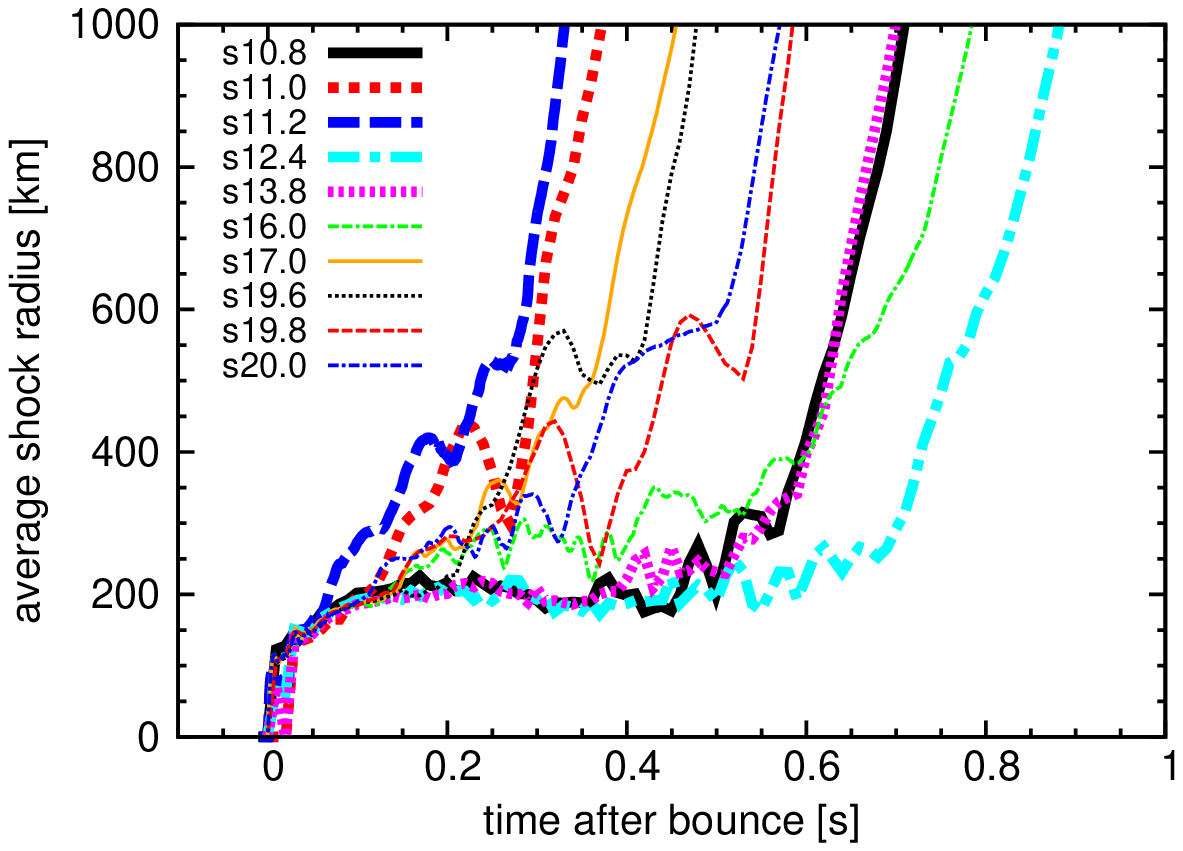} 
\end{tabular}
\end{center}
\caption{Similar to Figure \ref{f3} but for the (gravitational) NS mass (left panel) and the average shock radius (right panel). 
Here, gravitational NS mass is defined as the rest mass, 
$\int \rho \, dV$, 
minus the mass equivalent of the binding energy,
$\frac{1}{2} | \int \rho \Phi \, dV | $. 
Note in the right panel that only the post-bounce evolution up to $ 1$ s is shown to focus on the shock revival time.}
\label{f4}
\end{figure*}

From Equation (\ref{eq:3}), the mass of the (forming) NS, $M_{\rm NS}$, also affects the kick velocity ($v_{\rm NS}$). Figure \ref{f4} shows the evolution of the gravitational mass of the forming NS (left panel) with the average shock radius (right panel) in all the 2D models. 
By comparing the two panels, one can see that early shock revival (for example, model s11.2, thick blue dashed line) leads to the formation of a less massive NS. 
Note that the NS masses for models s10.8 (thick black solid line) and s12.4 (thick cyan dash-dotted line) are relatively small, although the shock revival time is late. This is because of the small mass accretion rate onto the PNS, which is characterized by the small progenitor's compactness parameter (see $\xi_{2.5}$ in Table \ref{t1}). The NS masses of our 2D models are in the range  $1.65 \pm 0.3 \Msun$, where the variation (at most $\sim 30 \%$) is much smaller than that of $\vns$. 
Our results suggest that the impact of $M_{\rm NS}$ on the 
kick velocity (see Equation (\ref{eq:3})) is weaker compared to that of $P_{\rm gas}$. 
Note again that $P_{\rm gas}$ is well correlated with the (diagnostic) explosion energy, as already mentioned. Therefore, small $M_{\rm NS}$ leads to small $\vns$ (not large $\vns$, although $M_{\rm NS}$ is in the denominator of Equation (\ref{eq:3})) because the small mass accretion rate to the PNS results in weaker explosion (via the small accretion neutrino luminosity), leading to a reduction in the total momentum of the ejecta ($P_{\rm gas}$). 

It should be noted that only three models (s10.8, s11.0, and s11.2), which have the smallest ZAMS mass and compactness among the examined models, leave a PNS with typical observational mass ($\sim 1.4 \Msun$). 
The other models present higher $M_{\rm NS}$ ($> 1.6 \Msun$) and it is still growing at the end of the simulations. 
Although standard initial mass functions (IMF), such as the Salpeter IMF, predict that heavy progenitors are subdominant, their too-massive PNS might be caused by some durable downflows peculiar to 2D axisymmetric simulations. 
This will be discussed in the Appendix \ref{sec:accr} in detail.

\subsection{Comparison with observation}
\label{sec:obs}

As mentioned in the previous section, the kick velocity is predominantly determined by $P_{\rm gas}$ in Equation (\ref{eq:3}), or most equivalently by the strength of the explosion. Since the high progenitor compactness leads to the energetic explosion via the high (accretion) neutrino luminosity \citep{nakamura15}, one could imagine that there is a possible correlation between the progenitor's compactness parameter and the kick velocity.

\begin{figure*}[htb]
\begin{center}
\begin{tabular}{cc}
\includegraphics[width=0.5\linewidth]{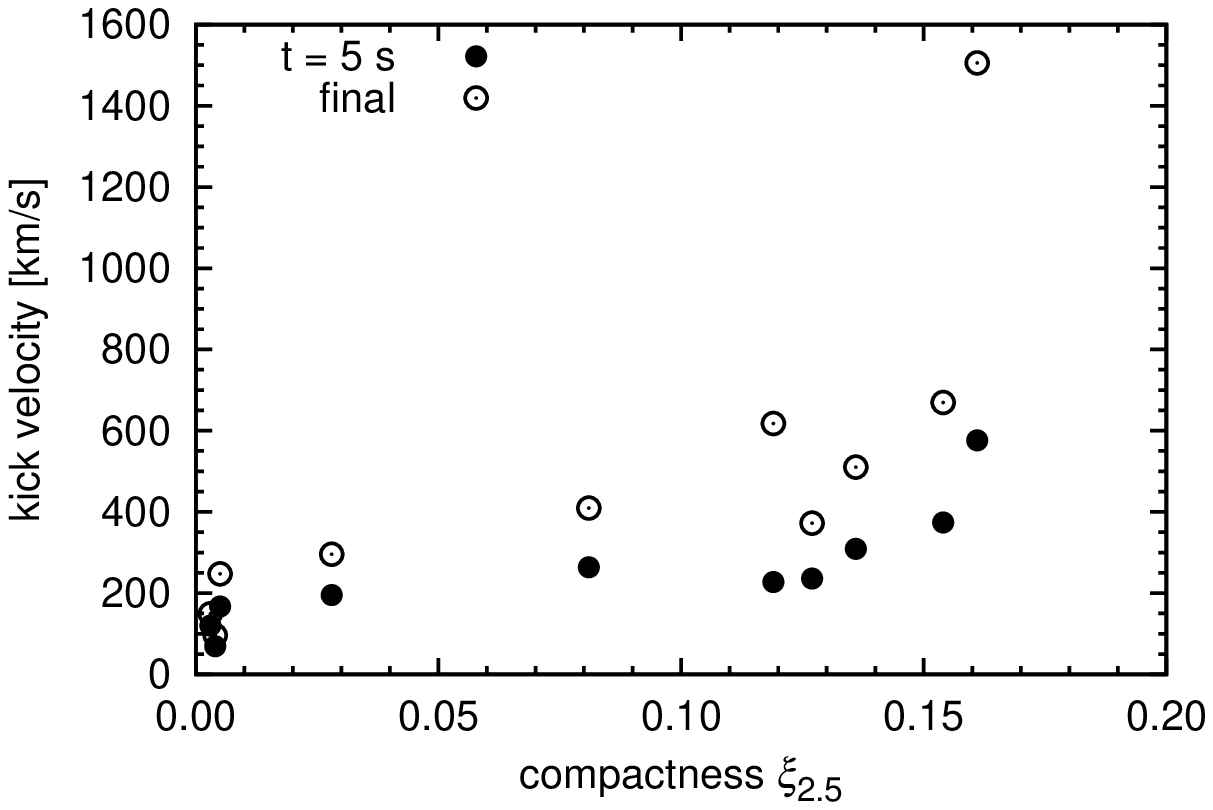} &
\includegraphics[width=0.5\linewidth]{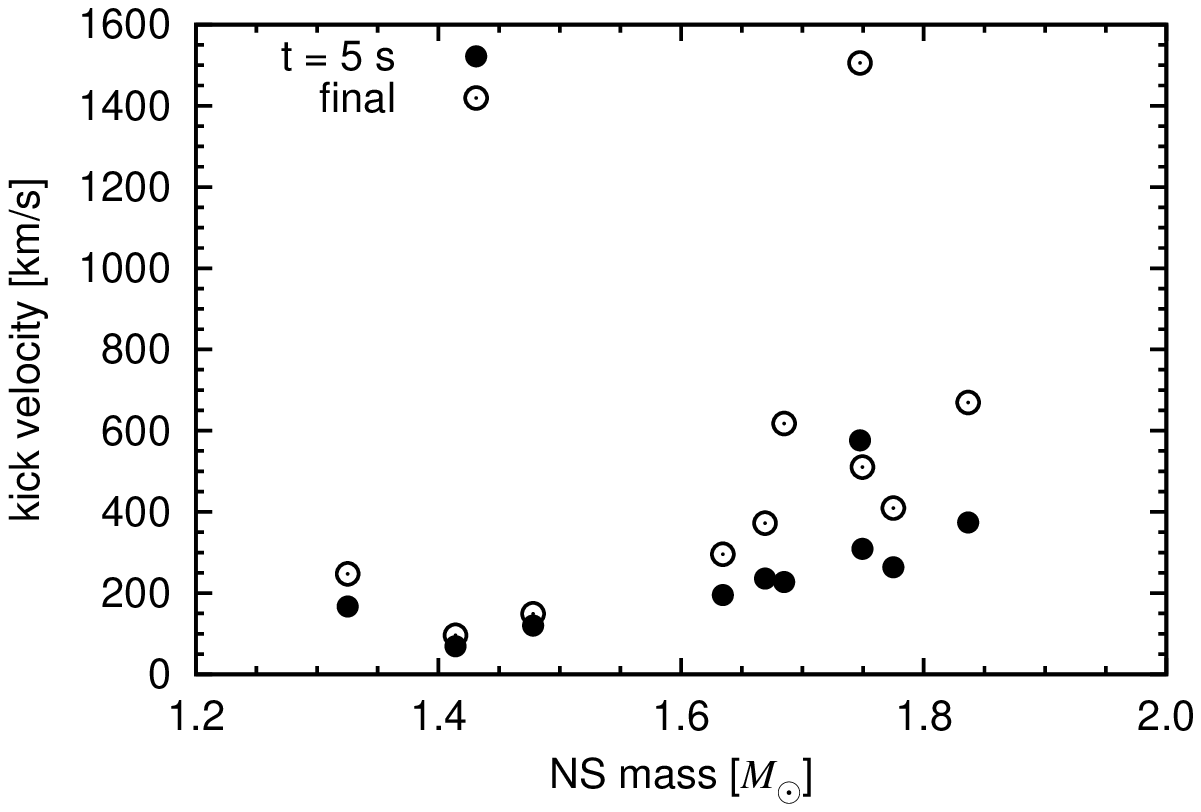}
\end{tabular}
\end{center}
\caption{NS kick velocity as a function of the compactness parameter $\xi_{2.5}$ (left panel) and gravitational NS mass (right panel). The kick velocities shown by filled circles, as well as NS mass in the right panel, are estimated at 5 s after bounce. The kick velocities shown by open circles are the estimated final values using Equation (\ref{eq:extra}).}
\label{f5}
\end{figure*}

The left panel of Figure \ref{f5} compares the kick velocities of our 2D models with the progenitor's compactness parameter $\xi_{2.5}$. 
We estimate the kick velocity at 5 s after bounce (filled circles), when the kick velocity is available for all the models given the different final simulation time. 
Also shown are the final values of the kick velocity (open circles) estimated by assuming ${\rm d} \vns /{\rm d}t = a \, t^{-2}$ \citep{bernhard19}, which gives the final kick velocity as
\begin{equation}\label{eq:extra}
    \vns(t) = -a \, t^{-1} + \vns^{\rm f}.
\end{equation}
The parameter $a$ and the final kick velocity $\vns^{\rm f}$ are determined by fitting the simulation data over an interval of 0--1 s before the end of each simulation. 
Both of them show a rough correlation to the compactness. 
It may not be surprising that the kick velocity is also correlated with $M_{\rm NS}$ (right panel of Figure \ref{f5}).

The progenitor's compactness parameter is, however, not directly observable. 
To get an idea of the relation between the NS mass and 
the kick velocity, we summarize in Table \ref{t2} the masses \citep[][and references therein]{antoniadis16} of millisecond pulsars 
in a binary system 
and the tangential velocity of the proper motions obtained by radio timing observations and optical spectroscopy  \citep{desvignes16,matthews16}. 
The table suggests that the more massive pulsars have a tendency to have a higher velocity. 
Readers should be aware that NSs in Table \ref{t2} had to maintain the binary system after the SN explosion they had once experienced. 
This means that their explosion was weak and/or the ejected mass was small, both of which could lead to smaller values of kick velocities compared with single NSs. 
Although more detailed analysis in order to clarify the relation between the observed proper motions and the NS natal kick is needed in order to draw a robust conclusion, we point out that the correlation found in Figure \ref{f5} is 
 compatible with the fact that the heaviest three NSs in Table \ref{t2} (J1909-3744, J0751+1807, and J1614-2230) have top-three kick velocities among the listed NSs. 
Here we do not include NS-NS binaries because they have undergone a second kick which obscures the possible correlation between NS kick and mass.

Table \ref{t3} summarizes some representative explosion properties of our 2D models such as the shock revival time, the diagnostic explosion energy, the mass of the PNS, the synthesized Ni mass, and the kick velocities. We define the shock revival time, $t_{500}$, as the time when an average shock radius reaches a radius of 500 km. Three kinds of kick velocities, $\vns$, $\vns^*$, and $\vns^{\rm f}$, are listed in the Table.  $\vns$ is from our simulations (Figures \ref{f1} and \ref{f5}), whereas $\vns^*$ is from a semi-analytical formula in \citet{janka17}, 
\begin{equation}
    \vns^* = 211 \, {\rm km \, s}^{-1} 
    \left( \frac{f_{\rm kin}}{\epsilon_5 \beta_\nu} \right)^{1/2}
    \left( \frac{\alpha_{\rm gas}}{0.1} \right)
    \left( \frac{E_{\rm dia}}{10^{51}{\rm erg}} \right)
    \left( \frac{M_{\rm NS}}{1.5 \Msun} \right)^{-1},
\end{equation}
where $\alpha_{\rm gas}$ and $M_{\rm NS}$ are the asymmetry parameter and baryonic NS mass as in Equation \ref{eq:3}, and $E_{\rm dia}$ is the diagnostic explosion energy. 
We assume the second factor on the right-hand side including $f_{\rm kin}$, the fraction of kinetic energy in the total explosion energy, to be unity (as suggested in \citet{janka17}) and the other values are extracted from our simulations. Although $\vns^*$ can be $\sim \, 30$ \% smaller than $\vns$, one can see that the analytical formula is 
able to quite nicely reproduce the main features of our results regarding the progenitor dependence of the kick velocities.

The values of the kick velocity in our 2D CCSN models range between 
69 and 576 km s$^{-1}$ at 5 s after bounce, 
$\sim 100$ and $\sim 850$ km s$^{-1}$ at the final time of the simulations, 
and $\sim 100$ and $\sim 1500$ km s$^{-1}$ in the estimated final values using Equation(\ref{eq:extra}). . 
Although this is, at least, compatible with the observed values (200--500 km s$^{-1}$ for the youngest pulsars and $\sim 1000$ km s$^{-1}$ for the fastest one), one should examine this in 3D models. However, 3D self-consistent, long-term CCSN simulations covering a wide range of progenitor mass (and compactness) are computationally very demanding. We shall limit ourselves to reporting two 3D runs, which we move on to explain in the next section.

\begin{table}[htb]
\caption{Pulsar properties}\label{t2}
\begin{center}
\begin{tabular}{lrr}
\hline 
Name & Mass ($\Msun$) & velocity (km s$^{-1}$)\\
\hline 
J1918-0642 & 1.18 & 43\\
J1802-2124 & 1.24 & 13\\
B1855+09    & 1.30 & 42\\
J1713+0747 & 1.31 & 37\\
J1738+0333 & 1.47 & 60\\
J1909-3744 & 1.54 & 200\\
J0751+1807 & 1.64 & 85\\
J1614-2230 & 1.93 & 110\\
\hline 
\end{tabular}
\end{center}
\end{table}

\begin{table*}[htb]
\caption{Overview of explosion properties obtained in our 2D models. Shock revival time ($t_{500}$), diagnostic explosion energy ($E_{\rm dia}$), gravitational NS mass (grav. $M_{\rm NS}$), ejected Ni mass ($M_{\rm Ni}$), and NS kick velocities ($\vns$, $\vns^*$) are estimated at 5 s after bounce. The final kick velocity ($\vns^{\rm f}$) is obtained from Equation (\ref{eq:extra}).
}\label{t3}
\begin{center}
\begin{tabular}{cccccccc}
\hline 
Progenitor & $t_{500}$ & $E_{\rm dia}$ & grav. $M_{\rm NS}$ & $M_{\rm Ni}$ & $\vns$ & $\vns^*$ & $\vns^{\rm f}$\\
                  & (s)            & (foe)              & ($\Msun$) & ($10^{-2} \Msun$) & (km s$^{-1}$) & (km s$^{-1}$) & (km s$^{-1}$)\\
\hline 
s10.8 & 0.619 & 0.125 & 1.48 & 0.482   & 120 & 79 & 149\\
s11.0 & 0.300 & 0.143 & 1.41 & 0.503   & 69  & 43 & 96\\
s11.2 & 0.236 & 0.173 & 1.33 & 0.499   & 167 & 105 & 247\\
s12.4 & 0.760 & 0.232 & 1.63 & 0.638   & 195 & 151 & 296\\
s13.8 & 0.630 & 0.303 & 1.77 & 0.905   & 263 & 190 & 409\\
s16.0 & 0.641 & 0.452 & 1.84 & 1.36     & 374 & 265 & 669\\
s17.0 & 0.362 & 0.867 & 1.75 & 2.32     & 576 & 486 & 1505\\
s19.6 & 0.300 & 0.265 & 1.68 & 0.629   & 228 & 184 & 618\\
s19.8 & 0.450 & 0.374 & 1.75 & 0.878   & 309 & 216 & 510\\
s20.0 & 0.388 & 0.399 & 1.67 & 1.14     & 236 & 198 & 372\\
\hline 
\end{tabular}
\end{center}
\end{table*}

\section{3D simulation of an $11.2 M_{\odot}$ star}
\label{sec:3d}

In this section we attempt to explore the long-term ($> 1$ s) evolution of a 3D model of a $11.2 M_{\odot}$ star. To make this doable, we relax the Courant-Friedrichs-Lewy (CFL) condition by means of the "mesh coarsening" technique, by which one can significantly reduce the computational cost.
In spherical coordinates, the CFL condition is quite severe around the polar
axis, especially close to the center since the width of the numerical grid $\Delta l$ ($= r \sin \frac{\theta}{2} \, {\rm d}\phi \sim \frac{1}{2}r \, {\rm d}\theta \, {\rm d}\phi$) becomes small
 there. To get around this problem, we combine the neighboring cells inside the PNS (``coarsen'' the numerical grid) and make the timestep longer there by evaluating the timestep over the combined cells divided by the velocities averaged over the original cells (see \citet{bernhard19} for a more sophisticated mesh-coarsening scheme). The mesh coarsening level is arbitrary.
We set the level to the maximum (1 zone in the $\theta$--$\phi$ direction or spherical) within 10 km and then downgrade step by step
($2 \times 4$ zones, $4 \times 8$ zones, to be in accordance with the original resolution). 
Apart form the use of mesh coarsening, the numerical code is the same for both 2D and 3D cases.

\begin{figure*}[ht]
\begin{center}
\begin{tabular}{cc}
\includegraphics[width=0.48\textwidth]{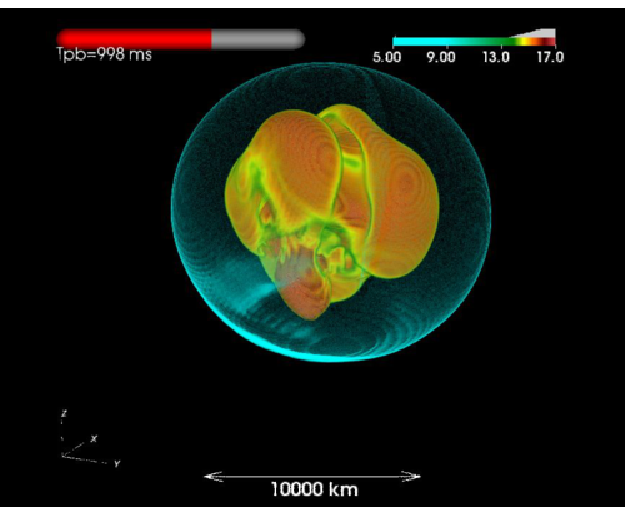} &
\includegraphics[width=0.48\textwidth]{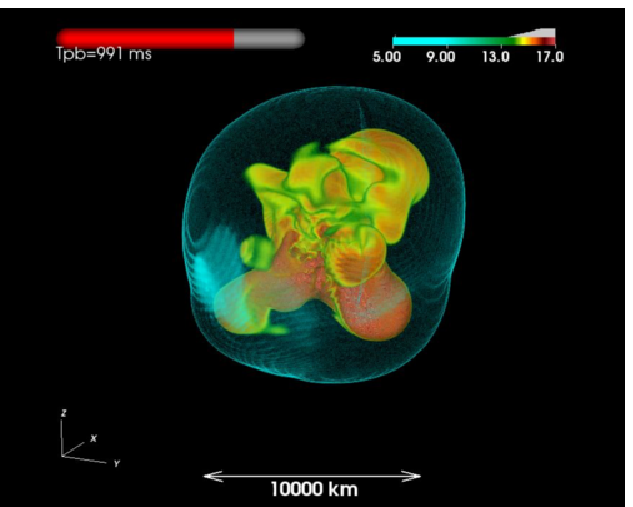}
\end{tabular}
\caption{3D entropy maps of the $11.2 \Msun$ model with (left panel) and without (right panel) the mesh coarsening scheme at $\sim 1$ s after bounce. The position of the shock is highlighted by a semitransparent surface colored by bright cyan for both of the models.}
\label{f6}
\end{center}
\end{figure*}

Figure \ref{f6} shows color-coded snapshots of the entropy for the 3D models of the 11.2 $M_{\odot}$ star with (left panel) and without (right panel) the mesh coarsening. 
It can be seen that the 3D models have a nearly spherical shock structure, 
although the shock expansion in the model using mesh coarsening (left panel) tends to be aligned with the polar axis (the $z$ axis in Figure \ref{f6}). 
On the other hand, the model without mesh coarsening 
has a random distribution of high entropy regions behind the shock. 
Its shock front is deformed at this time but does not have a specific orientation, in stark contrast to our 3D model using mesh coarsening.

\begin{figure*}[htb]
\begin{center}
\begin{tabular}{cc}
\includegraphics[width=0.48\textwidth]{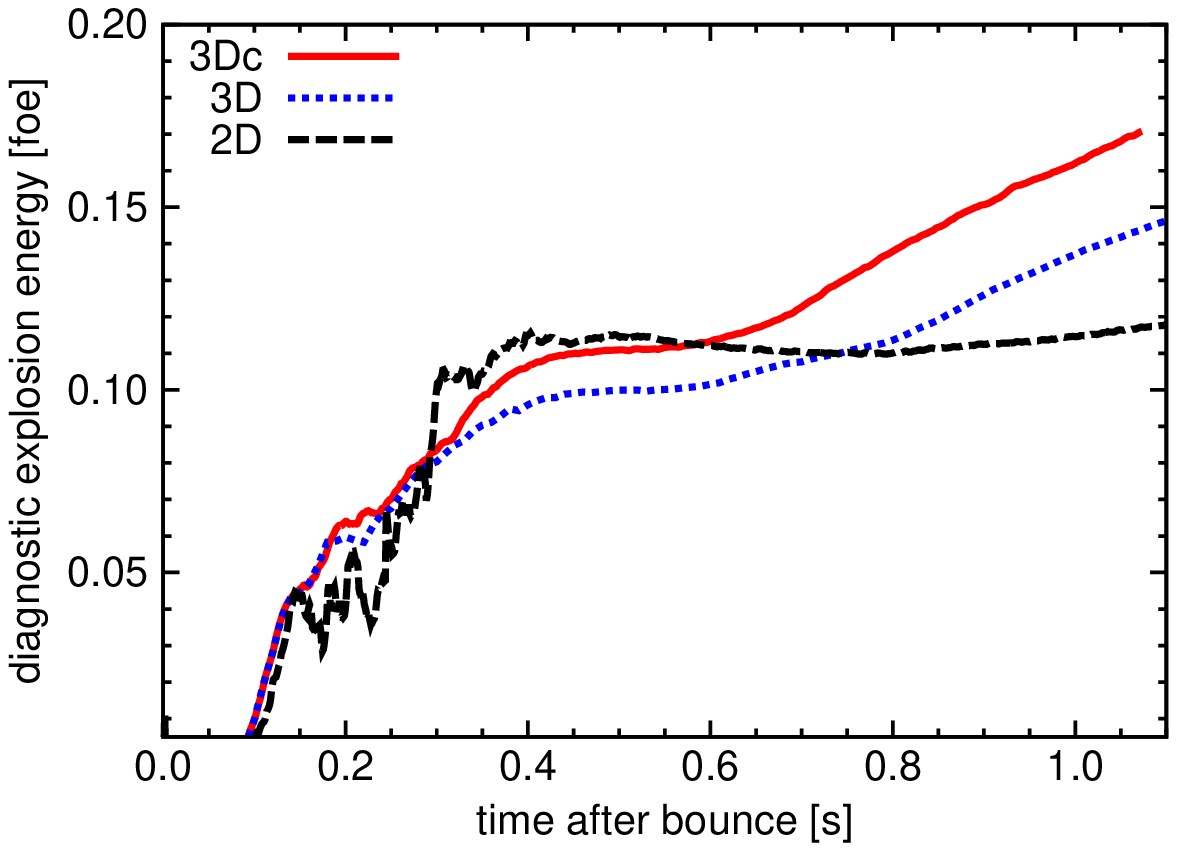} &
\includegraphics[width=0.48\textwidth]{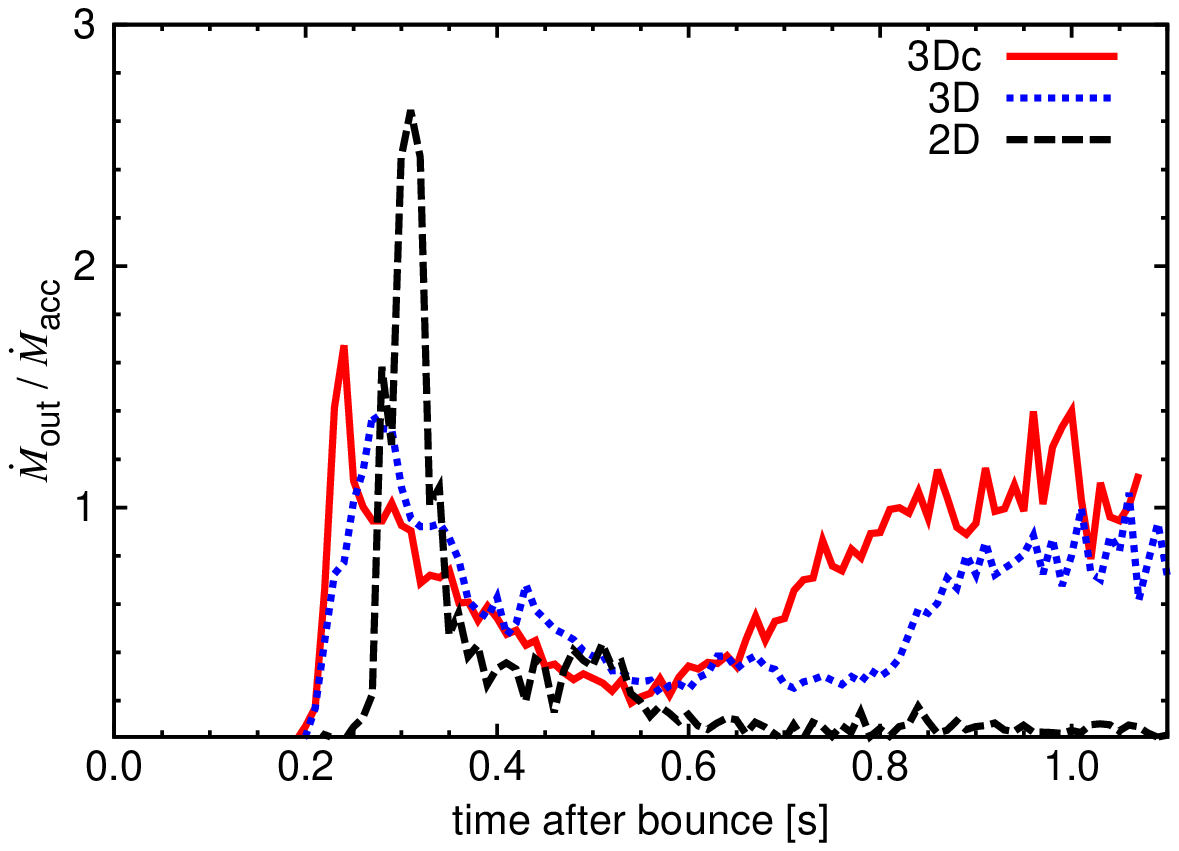} 
\end{tabular}
\caption{The diagnostic explosion energy (left panel) and the ratio of the mass outflow rate $\dot{M}_{\rm out}$ to the mass accretion rate $\dot{M}_{\rm acc}$ (right panel) as a function of the post-bounce time. The mass flow rate is measured at a radius of 500 km. Shown are 3D models with and without mesh coarsening (labeled as 3Dc (red solid line) and 3D (blue dashed line), respectively)  of the $11.2 \Msun$ star compared with the corresponding 2D model (black dashed line).}
\label{f7}
\end{center}
\end{figure*}

The left panel of Figure \ref{f7} compares the diagnostic explosion energy between the 3D models and the corresponding 2D model.
Our 3D models have higher explosion energy than the 2D model when their maximum shock radius reaches the outer boundary of the simulation regime (10,000 km, 1/10 of the 2D models) when the 3D runs are terminated.
This is consistent with the result of \citet{muellerb15}, who employed the same progenitor model.
The energetic explosion of the 3D models is driven by outflow from the central region.
The right panel of Figure \ref{f7} presents the ratio of the mass outflow rate $\dot{M}_{\rm out}$ to the mass accretion rate $\dot{M}_{\rm acc}$ measured at a radius of 500 km. 
The 3D models evolve in a similar way to the 2D model until $\sim 0.5$ s after bounce, then $\dot{M}_{\rm out}$ increases to become comparable to $\dot{M}_{\rm acc}$ and the explosion energy of the 3D models overwhelms that of the 2D model.
The flow ratio of the 2D model falls below unity at $\sim 0.35$ s and after that the explosion energy does not grow. 
The mass accretion rate of the 2D model becomes higher than the outflow rate, 
but small because of the small progenitor's compactness in model s11.2, and the PNS mass increases slowly
 with time (Figure \ref{f4}, left panel).

\begin{figure*}[htb]
\begin{center}
\begin{tabular}{cc}
\includegraphics[width=0.48\textwidth]{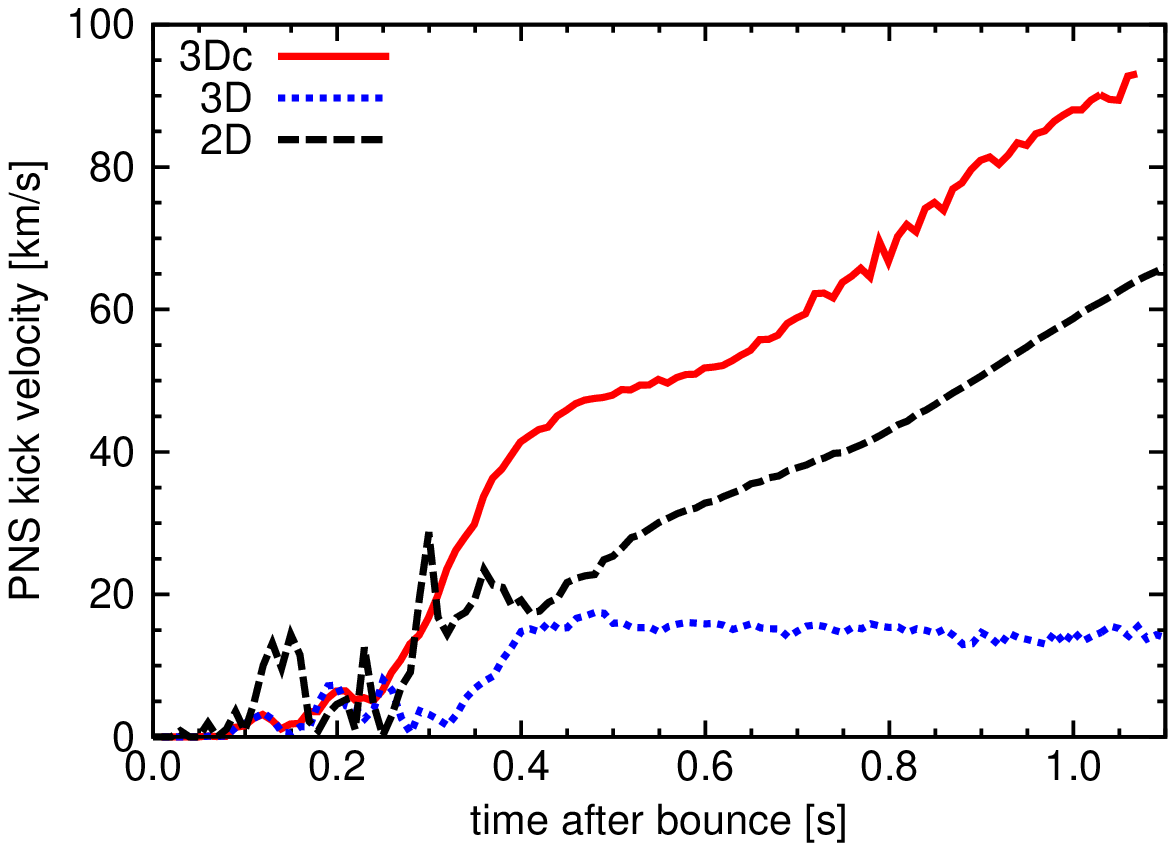} &
\includegraphics[width=0.48\textwidth]{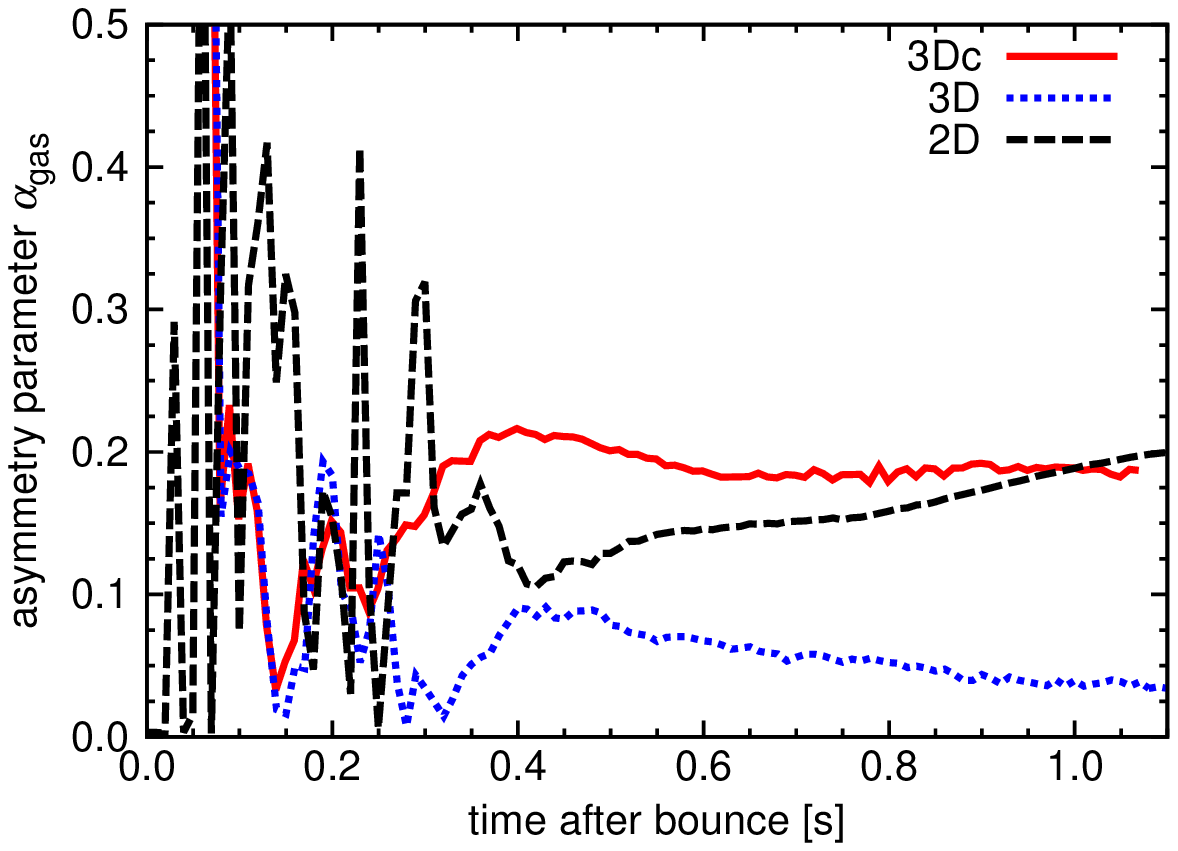} 
\end{tabular}
\caption{Same as Figure \ref{f7} but for the kick velocity (left panel) and the asymmetry parameter (right panel).}
\label{f8}
\end{center}
\end{figure*}

To estimate the kick properties from the 3D models, one needs to slightly modify equation (\ref{eq:2}) as $\alpha_{\rm gas} \equiv | \vec{P}_{\rm gas} | / P_{\rm gas} \equiv | \int {\rm d}m \,\vec{v} | /  \int {\rm d}m \,|\vec{v}|.$
Among the 3D models, 
the model simulated with mesh coarsening (model 3Dc) has an oriented shock expansion and a larger explosion energy than the 3D model without mesh coarsening (model 3D). 
This results in its asymmetry parameter being as high as the corresponding 2D model (Figure \ref{f8}, right panel) and the highest kick velocity among the models shown here (Figure \ref{f7}, left panel). 
The kick velocity of the 2D model is $66 \, {\rm km \, s}^{-1}$ at 1.1 s after bounce. 
The kick velocity of the 3Dc model at that time is $\sim 96 \, {\rm km \, s}^{-1}$, about 50 \% higher (left panel of Figure \ref{f8}). 
This is caused by a $\sim 50$ \% larger explosion energy in the 3Dc model ($1.7 \, \times \, 10^{50}$ erg) than in the 2D model ($1.2 \, \times \, 10^{50}$ erg). 
On the other hand, the kick velocity of the 3D model without the coarsening method is small ($14 \, {\rm km \, s}^{-1}$, 80 \% smaller than the 2D kick), which is caused by the nearly spherical distribution of the ejecta (small $\alpha_{\rm gas}$). 
Our results demonstrate that the use of mesh coarsening, albeit quite useful for making long-term 3D runs possible, could make the shock expansion align closely along the polar axis, possibly leading to an overestimation of the kick velocity. 
It should be noted that CCSN properties such as NS kick velocity could be affected by the stochastic nature of nonlinear hydrodynamics. It is still unclear whether the disagreement between the two 3D models is caused by the mesh coarsening or merely a result of the stochasticity. 
Examined here is only one progenitor with a ZAMS mass of $11.2 \, \Msun$, and the results might depend on the coarsening level as well as the progenitor structure. 
More detailed study is apparently needed to unambiguously pin down the impact of the mesh coarsening on the explosion properties in 3D CCSN models.

\section{Summary and Discussion}
\label{sec:discuss}

We have investigated the properties of the kick velocities of the forming NSs based on our long-term hydrodynamics CCSN simulations. 
We performed 2D simulations for ten progenitors from a 10.8 to 20 $\Msun$ star covering a wide range of progenitor compactness parameter, and two 3D runs for an 11.2 $\Msun$ star. 
Our 2D models presented a variety of the explosion energies between $\sim 1.3 \times 10^{50}$ erg and $\sim 1.2 \times 10^{51}$ erg, and NS kick velocities between $\sim 100$ km s$^{-1}$ and $\sim 1500$ km s$^{-1}$. 
For the 2D exploding models, 
it was found that the total momentum of the ejecta, or the explosion energy, is a predominant factor determining the kick velocity, whereas the ejecta asymmetry and NS mass play a secondary role. 
We also found that the kick velocities tend to become higher with the progenitor's
compactness. This is because high progenitor compactness results in high neutrino luminosity from the PNS, leading to more energetic explosions.
Since the high-compactness progenitors produce
massive PNS, we point out that the NS masses
and the kick velocities can be correlated
(very recently similar conclusion was obtained in \citet{bernhard19}, see their Figure 12 for detail), which we point out is
moderately supported by observation of pulsars in binary systems.
Comparing the 2D and 3D models of the 11.2 $\Msun$ star, 
the diagnostic explosion energy in 3D is, as 
previously identified, higher than that in 2D, whereas the 3D model results in smaller asymmetry in the ejecta distribution and 
smaller kick velocity than in 2D. The kick velocity of the 3D model is $\sim \, 80$ \% smaller than that of 2D model. We discussed some possible drawbacks of using the mesh coarsening to estimate the kick velocity.

Our CCSN models even in 2D do not reach typical SN observational values in terms of the smaller explosion energies (less than $\sim 10^{51}$ erg, except for model s17.0) and the level of the synthesized $^{56}$Ni. Some possible missing ingredients to make the underpowered explosion more energetic (also in 3D) should include multi-dimensional effects 
during the final stage of the pre-supernova evolution
(see \citet{couch_ott13,rodorigo14,couch_ott15,bernhard15,Burrows18,yoshida19}), general relativity (e.g., \citet{BMuller12a,Ott13,KurodaT12,KurodaT16}),
rapid rotation and/or magnetic fields (e.g., \citet{Marek09,Suwa10,takiwaki16,Summa18,harada18,martin06,moesta14,jerome15,masada15,martin17}), sophistication in the neutrino opacities \citep{melson15b,bollig17,Burrows18,kotake18} and the transport schemes (e.g., \citet{sumi12,richers17,nagakura18,just18}),
 and possibly inclusion of the quark-hadron phase transition in the center of the PNS \citep{tobias18}. 
As for predicting NS kicks, systematic study based on 3D CCSN modeling including a suite of the above missing facets is mandatory for making quantitative CCSN multi-messenger predictions possible. This study is nothing but a very first step toward the final goal.

\begin{ack}
We thank Masaomi Tanaka and Masaki Yamaguchi for  helpful discussions. 
This study was supported in part by Grants-in-Aid for Scientific Research of the Japan Society for the Promotion of Science (JSPS, Nos. 
JP26707013, 
JP26870823, 
JP16K17668, 
JP17H01130, 
JP17K14306, 
JP18H01212), 
the Ministry of Education, Science and Culture of Japan (MEXT, Nos. 
JP15H00789, 
JP15H01039, 
JP15KK0173, 
JP17H05205, 
JP17H05206, 
JP17H06357, 
JP17H06364, 
JP17H06365, 
JP24103001 
JP24103006, 
JP26104001, 
JP26104007), 
 by the Central Research Institute of Explosive Stellar Phenomena (REISEP) at Fukuoka University and associated projects (Nos.171042,177103),
and JICFuS as a priority issue to be tackled by using the Post ``K'' Computer. 
This work was also supported by the NINS program for cross-disciplinary study (Grant Numbers 01321802 and 01311904) on Turbulence, Transport, and Heating Dynamics in Laboratory and Solar/Astrophysical Plasmas: ``SoLaBo-X''.
\end{ack}

\bibliographystyle{apj}
\bibliography{reference}

\clearpage

\appendix

\section{Continuous Accretion in 2D models}
\label{sec:accr}
In this Appendix we revisit a caveat for the 2D models with high progenitor compactness, in which a downflow to the PNS is liable to continue for a long time, making the PNS too massive to be compared with observed ones (see also the detailed analysis by \citet{BMuller12a}.) 

The 2D models with smaller progenitor compactness
 (such as model 11.2 with $\xi_{2.5} = 0.005$, see Table \ref{t1}) can get around 
 this problem because of the small mass accretion and 
 the early shock revival. 
Models s10.8 and s11.0 also have very small compactness compared to the others, and their PNS mass is also small (1.48 and $1.41 \, \Msun$). 
The diagnostic explosion energy is also small for these three models (1.25 -- $1.73 \times 10^{50}$ erg). 
These results are consistent with the previous long-term 2D CCSN simulation by \citet{muellerb15}, 
where 2D simulations for progenitors in a similar mass range ($11.0 \Msun$ -- $11.6 \Msun$) results in $1.3 \Msun$ -- $1.6 \Msun$ PNS (baryonic) mass and 0.50 -- 2.1 $\times 10^{50}$ erg diagnostic energy.

\begin{figure}[ht]
\begin{center}
\includegraphics[width=0.48\textwidth]{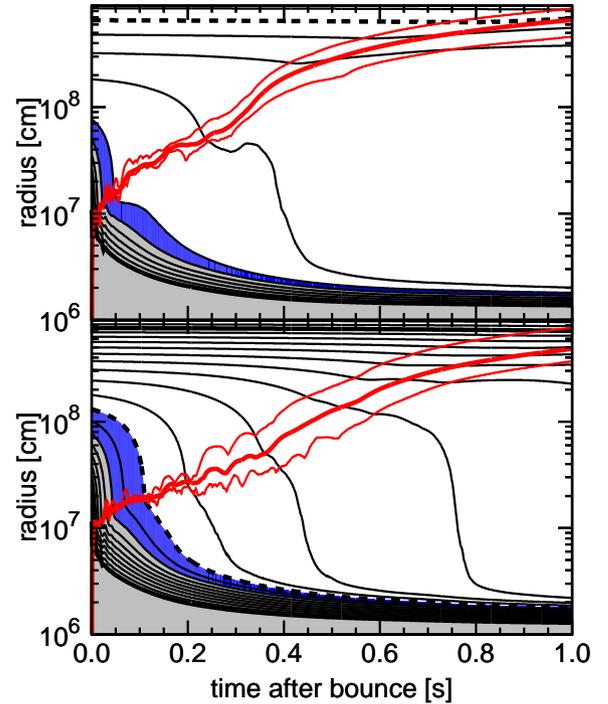} 
\caption{Mass shell diagram for models s11.2 (top panel) and s17.0 (bottom panel). The thick black lines correspond to the mass coordinate at every $0.5 \, \Msun$, with the thin lines at every $0.05 \, \Msun$. The dashed lines show the mass coordinate of $1.5 \, \Msun$ for reference. The color-shaded regions present the iron core (grey) and the silicon layer (blue) of each progenitor. The shock positions (red lines: maximum, average, and minimum from top to bottom) turn to go outward when the Si/SiO interface falls onto the shock. }
\label{f:a1}
\end{center}
\end{figure}

For the 2D models with higher progenitor compactness, the PNS (gravitational) masses keep growing during the simulation and finally become higher than typical observational value 
($\sim 1.4 \Msun$). The PNSs of these models are exposed to continuous mass accretion even after shock revival. 
This can be seen from Figure \ref{f:a1}, where the mass shell diagram for model s11.2 is compared with that of s17.0. 
These models show shock revival when the Si/SiO interface falls onto the shock. 
After the shock revival, the mass shells of model s11.2 is shown to turn to going outward and the mass accretion onto the PNS nearly stops.
On the other hand, the mass accretion to the PNS continues even after the shock revival for model s17.0 (bottom panel). 

\begin{figure*}
\begin{center}
\begin{tabular}{ccc}
\includegraphics[width=0.3\linewidth]{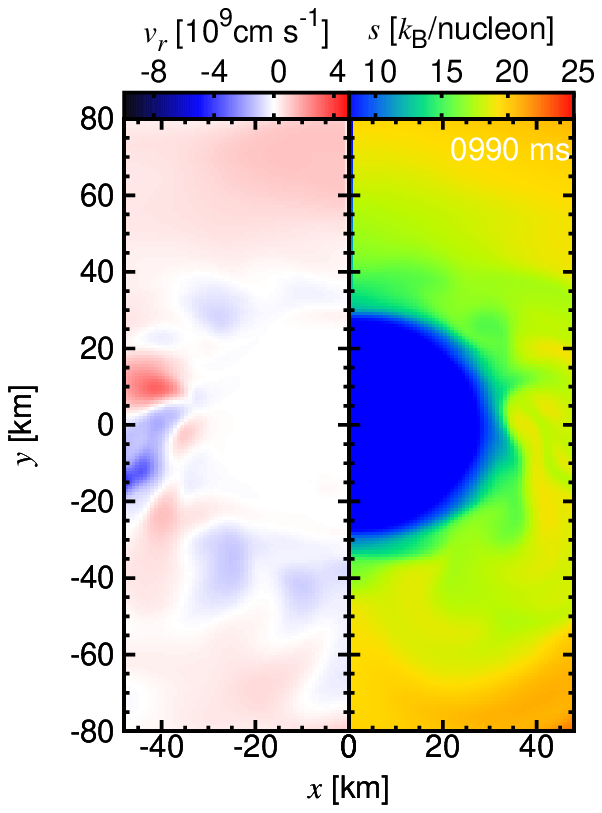} &
\includegraphics[width=0.3\linewidth]{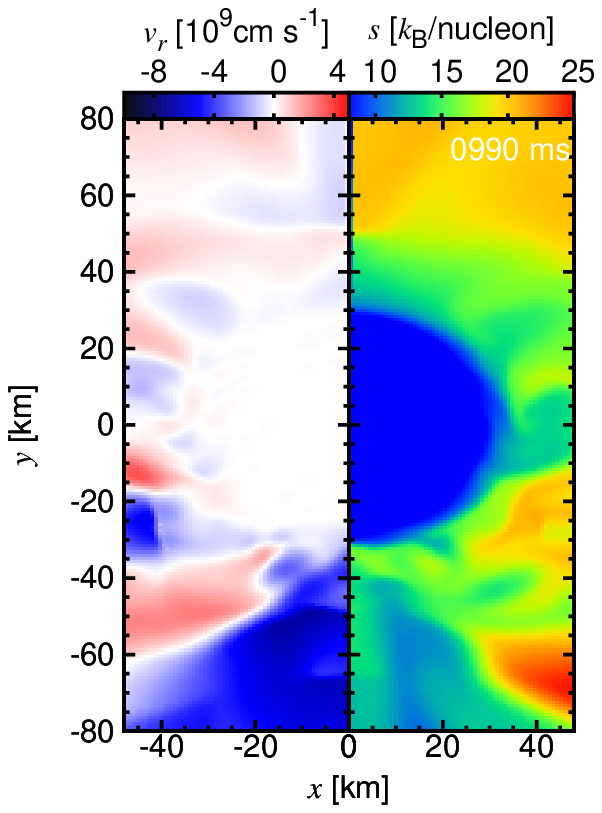} &
\includegraphics[width=0.3\linewidth]{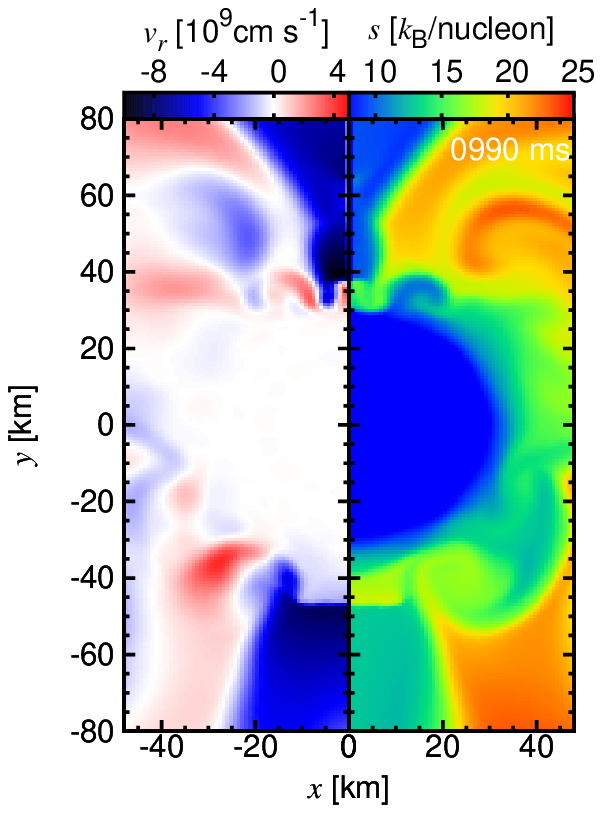} \\
\includegraphics[width=0.3\linewidth]{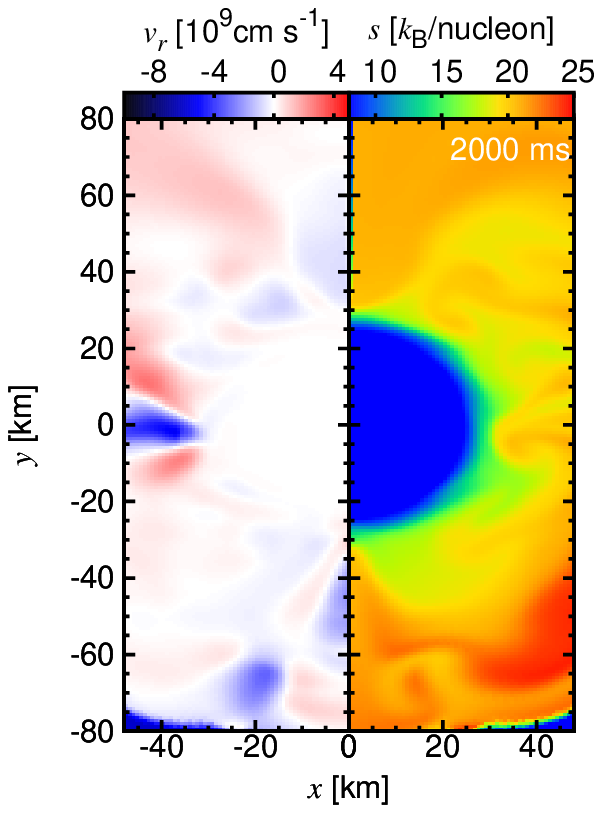} &
\includegraphics[width=0.3\linewidth]{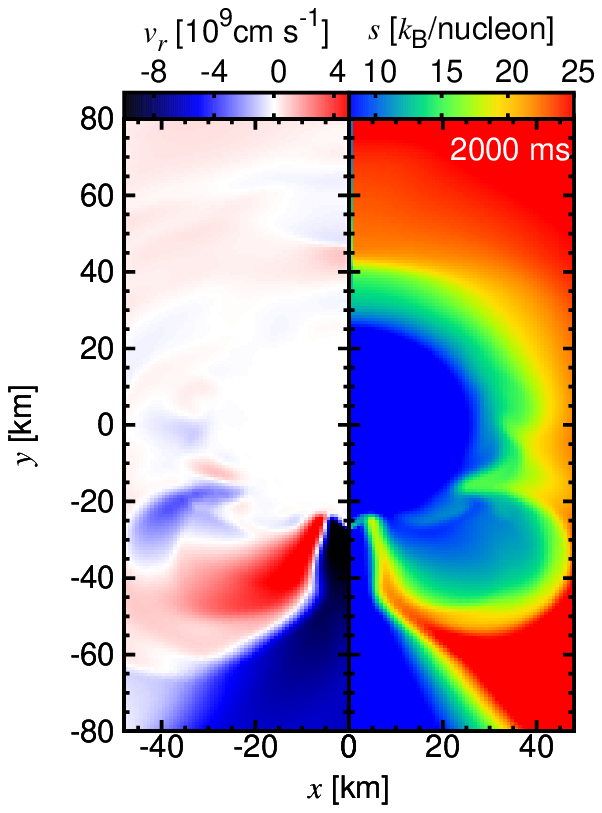} &
\includegraphics[width=0.3\linewidth]{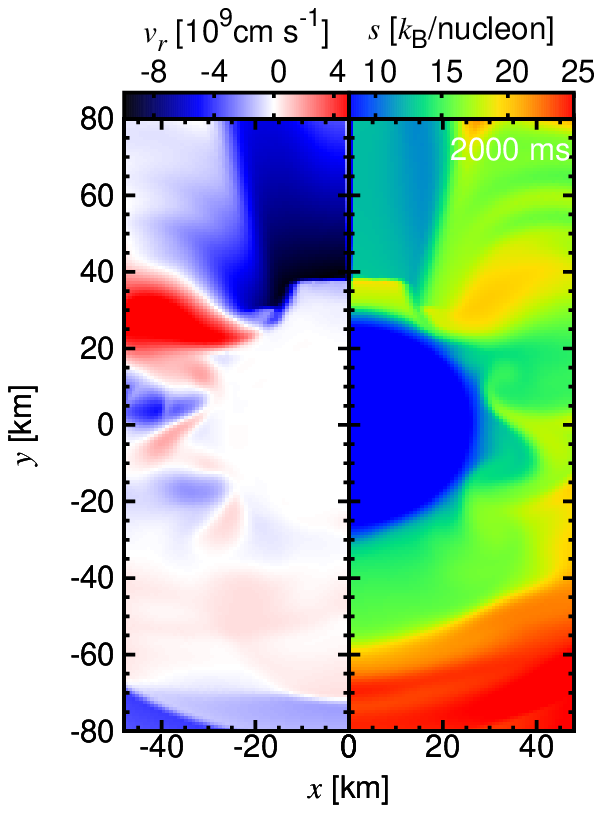} \\
\includegraphics[width=0.3\linewidth]{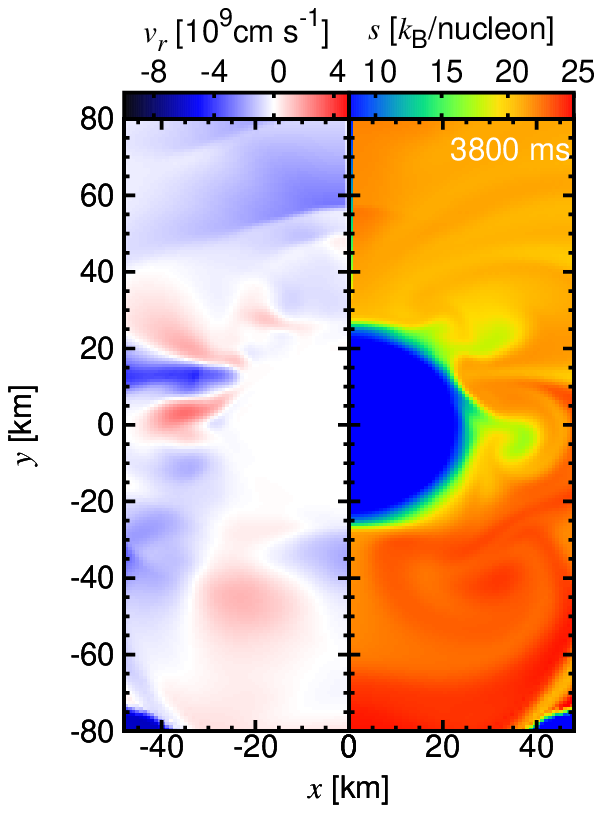} &
\includegraphics[width=0.3\linewidth]{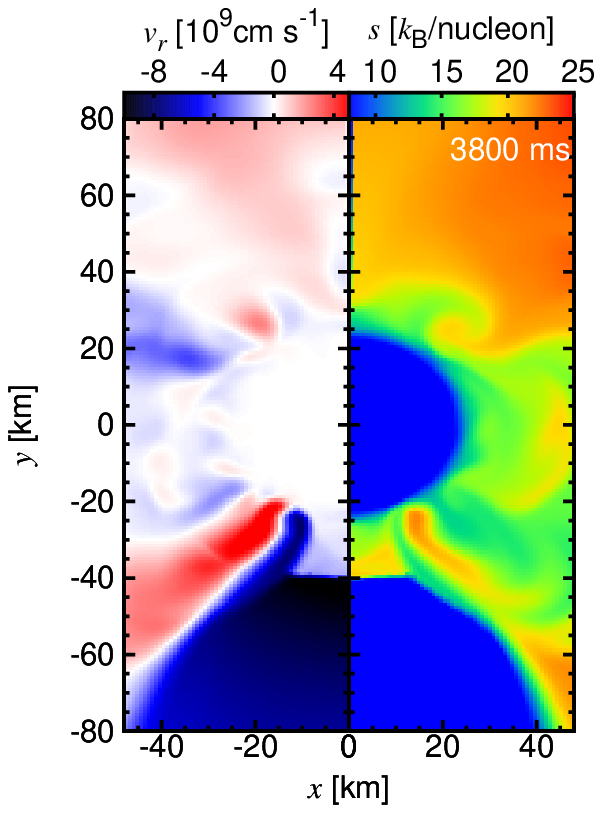} &
\includegraphics[width=0.3\linewidth]{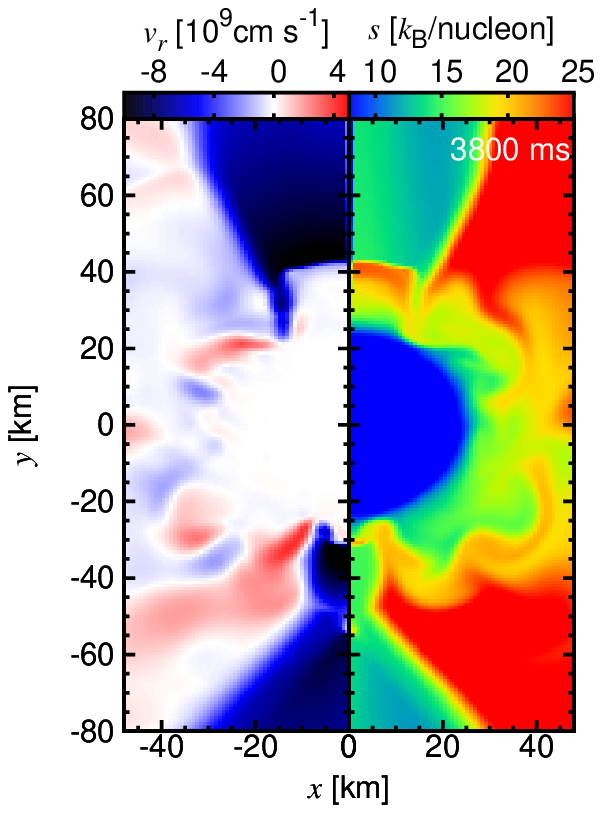} \\
\end{tabular}
\end{center}
\caption{Snapshots of the radial velocity $v_r$ (left column of each panel, in $10^9$ cm s$^{-1}$) and the specific entropy $s$ (right column, in $k_{\rm B}$/nucleon) for models s11.2 (left panels), s17.0 (middle panels), and s27.0 (right panels). Three time steps at post-bounce times of 990 ms, 2000 ms, and 3800 ms are shown from top to bottom. Note the absence of the low-entropy downflow onto the surface of the PNS for model s11.2 during a long-term simulation.}
\label{fa1}
\end{figure*}

To visualize this, we plot in Figure \ref{fa1} snapshots of the radial velocity and the entropy in the central regions of these two models and of an additional model, s27.0. 
The central spherical regions with low entropy and nearly zero radial velocity correspond to the PNS.
The PNS of model s11.2 (left panels) is surrounded by high-entropy gas heated by neutrinos. 
No clear long-lasting downflows to the PNS are observed. 
On the other hand, several strong downflows to the 
PNSs (colored by blue in the velocity plots) are clearly seen for models s17.0 (middle panels) and s27.0 (right panels). This feature is common to other 2D models with relatively high progenitor compactness. 
The infall flow wriggles around the PNS and, once it collides with its mirror flow on the symmetry axis, strong and durable down-flows from north and/or south poles to PNS are produced. 
Model s27.0, which has higher compactness ($\xi_{2.5}=0.232$) than the models examined in this paper, leaves behind a central remnant with a gravitational mass of $2.27 \Msun$ at the end of the simulation. 
This value is higher than the maximum mass of a cold NS ($2.04 \, \Msun$) for the currently employed LS220 EOS, although thermal pressure can leverage the maximum PNS mass. 

To assess the fate of this heavy remnant, we refer to 1D general relativistic simulations by \citet{oconnor11} using the same EOS. 
A linear fit to their results gives the maximum PNS mass as a function of the compactness \citep{nakamura15},
\begin{equation}
M_{\rm PNS,max}/\Msun = 0.52 \, \xi_{2.5} + 2.01.
\end{equation}
This formula gives $M_{\rm PNS,max} = 2.13 \Msun$ for model s27.0, and implies black hole (BH) formation at 5.28 s, although our Newtonian simulation does not have the ability to follow the BH formation.

\end{document}